\documentclass[9pt]{article}
\usepackage{extsizes}
\usepackage[super,sort&compress,comma]{natbib} 
\usepackage[version=3]{mhchem}
\usepackage[left=1.5cm, right=1.5cm, top=1.785cm, bottom=2.0cm]{geometry}
\usepackage{balance}
\usepackage{mathptmx}
\usepackage{sectsty}
\usepackage{graphicx} 
\usepackage{lastpage}
\usepackage[format=plain,justification=justified,singlelinecheck=false,font={stretch=1.125,small,sf},labelfont=bf,labelsep=space]{caption}
\usepackage{float}
\usepackage{fancyhdr}
\usepackage{fnpos}
\usepackage{orcidlink}
\usepackage[english]{babel}
\addto{\captionsenglish}{%
  
}
\usepackage{array}
\usepackage{droidsans}
\usepackage{charter}
\usepackage[T1]{fontenc}

\usepackage{setspace}
\usepackage[compact]{titlesec}
\usepackage{hyperref}

\usepackage{epstopdf}
\definecolor{cream}{RGB}{222,217,201}

\begin{document}

\pagestyle{fancy}
\thispagestyle{plain}
\fancypagestyle{plain}{
\renewcommand{\headrulewidth}{0pt}
}

\makeFNbottom
\makeatletter
\renewcommand\LARGE{\@setfontsize\LARGE{15pt}{17}}
\renewcommand\Large{\@setfontsize\Large{12pt}{14}}
\renewcommand\large{\@setfontsize\large{10pt}{12}}
\renewcommand\footnotesize{\@setfontsize\footnotesize{7pt}{10}}
\renewcommand\scriptsize{\@setfontsize\scriptsize{7pt}{7}}
\makeatother

\renewcommand{\thefootnote}{\fnsymbol{footnote}}
\renewcommand\footnoterule{\vspace*{1pt}%
\color{cream}\hrule width 3.5in height 0.4pt \color{black} \vspace*{5pt}} 
\setcounter{secnumdepth}{5}

\makeatletter 
\renewcommand\@biblabel[1]{#1}          \newcommand\nsurfaces{607 }
\newcommand\nasj{593 }
\newcommand\iuband{183921 }  
\renewcommand\@makefntext[1]%
{\noindent\makebox[0pt][r]{\@thefnmark\,}#1}
\makeatother 
\renewcommand{\figurename}{\small{Fig.}~}
\sectionfont{\sffamily\Large}
\subsectionfont{\normalsize}
\subsubsectionfont{\bf}
\setstretch{1.125} 
\setlength{\skip\footins}{0.8cm}
\setlength{\footnotesep}{0.25cm}
\setlength{\jot}{10pt}
\titlespacing*{\section}{0pt}{4pt}{4pt}
\titlespacing*{\subsection}{0pt}{15pt}{1pt}

\fancyfoot{}
\fancyfoot[LO,RE]{\vspace{-7.1pt}\includegraphics[height=9pt]{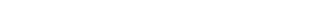}}
\fancyfoot[CO]{\vspace{-7.1pt}\hspace{13.2cm}\includegraphics{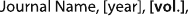}}
\fancyfoot[CE]{\vspace{-7.2pt}\hspace{-14.2cm}\includegraphics{head_foot/RF}}
\fancyfoot[RO]{\footnotesize{\sffamily{1--\pageref{LastPage} ~\textbar  \hspace{2pt}\thepage}}}
\fancyfoot[LE]{\footnotesize{\sffamily{\thepage~\textbar\hspace{3.45cm} 1--\pageref{LastPage}}}}
\fancyhead{}
\renewcommand{\headrulewidth}{0pt} 
\renewcommand{\footrulewidth}{0pt}
\setlength{\arrayrulewidth}{1pt}
\setlength{\columnsep}{6.5mm}
\setlength\bibsep{1pt}

\makeatletter 
\newlength{\figrulesep} 
\setlength{\figrulesep}{0.5\textfloatsep} 

\newcommand{\topfigrule}{\vspace*{-1pt}%
\noindent{\color{cream}\rule[-\figrulesep]{\columnwidth}{1.5pt}} }

\newcommand{\botfigrule}{\vspace*{-2pt}%
\noindent{\color{cream}\rule[\figrulesep]{\columnwidth}{1.5pt}} }

\newcommand{\dblfigrule}{\vspace*{-1pt}%
\noindent{\color{cream}\rule[-\figrulesep]{\textwidth}{1.5pt}} }

\makeatother

\sffamily
\begin{tabular}{m{4.5cm} p{13.5cm} }

\includegraphics{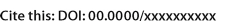} & \noindent\LARGE{\textbf{InterMat: Accelerating Band Offset Prediction in Semiconductor Interfaces with DFT and Deep Learning}} \\
 & \vspace{0.3cm} \\

 & \noindent\large{Kamal Choudhary\orcidlink{0000-0001-9737-8074}, Kevin F. Garrity\orcidlink{0000-0003-0263-4157} } \\
\includegraphics{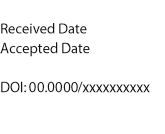} & \\

\end{tabular}


\renewcommand*\rmdefault{bch}\normalfont\upshape
\rmfamily
\section*{}
\vspace{-1cm}


\footnotetext{\textit{Material Measurement Laboratory, National Institute of Standards and Technology, Maryland, 20899, USA. E-mail: kamal.choudhary@nist.gov}}





\sffamily{\textbf{We introduce a computational framework (InterMat) to predict band offsets of semiconductor interfaces using density functional theory (DFT) and graph neural networks (GNN). As a first step, we benchmark OptB88vdW generalized gradient approximation (GGA) work functions and electron affinities for surfaces against experimental data with accuracies of 0.29 eV and 0.39 eV, respectively. Similarly, we evaluate band offset values using independent unit (IU) and alternate slab junction (ASJ) models leading to accuracies of 0.45 eV and 0.22 eV, respectively. We use bulk band structure calculations with the TBmBJ meta-GGA functional to correct for band gap underestimation when predicting conduction band properties. During ASJ structure generation, we use Zur's algorithm along with a unified GNN force-field to tackle the conformation challenges of interface design. At present, we have \nsurfaces surface work functions calculated with DFT, from which we can compute \iuband IU band offsets as well as \nasj directly calculated ASJ band offsets. Finally, as the space of all possible heterojunctions is too large to simulate with DFT, we develop generalized GNN models to quickly predict bulk band edges with an accuracy of 0.26 eV. We show how these models can be used to predict relevant quantities including ionization potentials, electron affinities, and IU-based band offsets. We establish simple rules using the above models to pre-screen potential semiconductor devices from a vast pool of nearly 1.4 trillion candidate interfaces. InterMat is available at website: \url{https://github.com/usnistgov/intermat}}}\\


\rmfamily 


\section*{Introduction}


Interfaces are critical for a variety of technological applications including semiconductor transistors and diodes, solid-state lighting devices, solar-cells, data-storage and battery applications \cite{butler2019designing,sutton1995interfaces,kroemer1982heterostructure,monch2013electronic,edward1992band,robertson2013band,smeu2021electron,agostini2011characterization}. In particular, the continued scaling of semiconductor devices towards the atomic limit \cite{taur1999incredible} makes interface properties even more important and a focus area of recent investments in research and development including the Creating Helpful Incentives to Produce Semiconductors (CHIPS) act\cite{nistCHIPSGov}. While interfaces are ubiquitous, predicting even basic interface properties from bulk data or chemical models remains challenging.  There have been numerous scientific efforts to model interfaces with a variety of techniques including density functional theory (DFT)\cite{van1985theoretical,franciosi1996heterojunction,weston2018accurate,hinuma2014band,ghosh2022efficient,di2021band,dardzinski2022best,mathew2016mpinterfaces,choudhary2023efficient,restuccia2023high}, force-field\cite{choudhary2015charge,yu2007charge,shan2010charge,hahn2015thermal}, tight-binding \cite{harrison1986tight,bernstein1998amorphous,schenter1992semiconductor,munoz1987tight} and machine learning techniques \cite{choudhary2023efficient,willhelm2022predicting,huang2019band,zhu2020fundamental}. However, to the best of our knowledge, there is no systematic investigation of interfaces for a large class of structural variety and chemical compositions. Most of the previous efforts focus on a limited number of interfaces, and hence there is a need for a dedicated infrastructure for data-driven interface materials design.




Some of the key quantities for determining interface properties are: equilibrium geometries, energetics, work functions, ionization potentials, electron affinities, band offsets, carrier effective masses, mobilities, and thermal conductivities. 
Calculations of band offsets and band-alignment at semiconductor heterojunctions are of special interest for device design. Semiconductor device transport and performance depend critically on valence band offsets ($\Delta E_v$) and conduction band offsets  ($\Delta E_c$), as well as interfacial roughness and defects \cite{milnes2012heterojunctions,robertson2005high,roul2015binary}. Based on the band-alignment, heterostructures can be categorized into three classes: (i) type-I (straddling gap), (ii) type-II (staggered gap), and (iii) type-III (broken gap). The type-I heterostructures are used for transistors, lasers and light-emitting diode (LED) applications, type-II are used for photoabsorbers and photocatalysts, and type-III are used for tunneling field effect transistors. 

Experimentally, band offsets can be measured using optical spectroscopy, X-ray photoelectron spectroscopy (XPS), ultraviolet photoelectron spectroscopy (UPS), and electrical measurements \cite{edward1992band}. However, these experiments can be quite time and resource consuming. Additionally, the variability across multiple reported measurements can be reasonably high. For example, the reported AlN/GaN interface $\Delta E_v$ varies from 0.57 eV to 1.36 eV with reported uncertainties of up to 0.24 eV\cite{roul2015binary}. In this respect, the computation of band offsets can serve as a complementary tool to experimental analysis. Nevertheless, the calculation of band offsets is rather challenging \cite{ohler1998heterojunction} and has been an area of research for about a century \cite{schottky1926small,bardeen1947surface,oliveira1985fano}. Density functional theory (DFT) calculations are one of the most widely used techniques for predicting band offsets, as they can describe the electronic and atomic structures at the interface in a self-consistent manner. There are two main approaches to predicting band offsets using DFT. The first is to directly simulate the interface using either an alternating slab-junction (ASJ)/superlattice or surface terminated junction (STJ)/slab vacuum geometry, either of which requires a computationally expensive calculation for each pair of materials. Alternatively, the independent unit (IU)/electron affinity/Anderson's model \cite{di2021band,anderson1960germanium} requires only independent surface calculations of each material, greatly reducing computational cost but ignoring specific interface effects. ASJ models were shown to be most accurate in Ref. \cite{di2021band}, but IU models are surprisingly competitive.

Importantly, the generation of an atomistic interface geometry is a challenging task due to the high number of possible conformations and configurations. There are several important factors determining an interface such as: the selection of the lattice alignment, the relative orientation/displacement between surfaces, the separation distance, point/line defects, and the presence of interfacial charge transfer. Several previous tools have attempted to address this challenge, including MPInterfaces \cite{mathew2016mpinterfaces}, TribChem \cite{losi2023tribchem} and QuantumATK \cite{smidstrup2019quantumatk} packages.

Moreover, DFT calculations of interfaces require initial pre-relaxed bulk structures which in this work are obtained from the Joint Automated Repository for Various Integrated Simulations (JARVIS)-DFT \cite{winesreview,choudhary2020joint} database containing nearly 80000 bulk 3D and 1100 2D materials. The JARVIS-DFT originated about 5 years ago and contains millions of properties materials and has carefully converged atomic structures with tight convergence parameters, various exchange-correlation functionals such as OptB88vdW\cite{klimevs2009chemical}, TBmBJ\cite{tran2009accurate}, R2SCAN \cite{furness2020accurate} and HSE06 \cite{heyd2003hybrid}. JARVIS-DFT contains metallic, semiconducting, insulator, superconductor , high-strength, topological, solar, thermoelectric, piezoelectric , dielectric, two-dimensional, magnetic, porous, defect and various other classes of bulk materials \cite{choudhary2022designing,wines2023high}. We have also previously looked at the band alignment of layered two dimensional materials using JARVIS-DFT \cite{choudhary2023efficient}. However, three dimensional systems with chemical bonding between the materials require much greater effort, as the interfacial bonding has a much greater effect on the interface properties, and the determination of even a single interface structure is a challenging task. Out of the above material class combinations, semiconductor-semiconductor are of special interest for this work.
As DFT calculations can be time-consuming for surfaces and interfaces machine-learning (ML)/deep learning (DL)  techniques based on DFT data can be used to accelerate atomistic predictions \cite{choudhary2022recent,vasudevan2019materials}. Such models have often been applied for bulk property predictions and their applicability for defects and interfaces remains an open question. Several machine learning tools available in JARVIS such as classical force-field inspired descriptors (CFID) \cite{choudhary2020joint}, atomistic line graph neural network (ALIGNN) \cite{choudhary2021atomistic,choudhary2023unified}, computer vision for atomistic images (AtomVision) \cite{choudhary2023atomvision} and natural language processing for chemistry (ChemNLP) \cite{choudhary2023chemnlp} can be used in this regards to accelerate the interface design tasks. In particular, ALIGNN has been used to develop several fast surrogate models for property predictions as well as a unified force-field for fast structure optimizations.

Most importantly, for all the above predictions, it is important to benchmark and quantify error with respect to experimental data to gain confidence in the prediction methodology. This work addresses the above challenges and provides a streamlined framework for semiconductor interface design (InterMat). Although focusing on semiconductors, this work has relevance to other applications such as battery, data-storage, and solar-cell devices. We believe that this work will be a precursor to more thorough theoretical and experimental investigations of semiconductor interfaces.

\section*{Results and discussion}
\begin{figure}[hbt!]
\centering
\includegraphics[width=\linewidth]{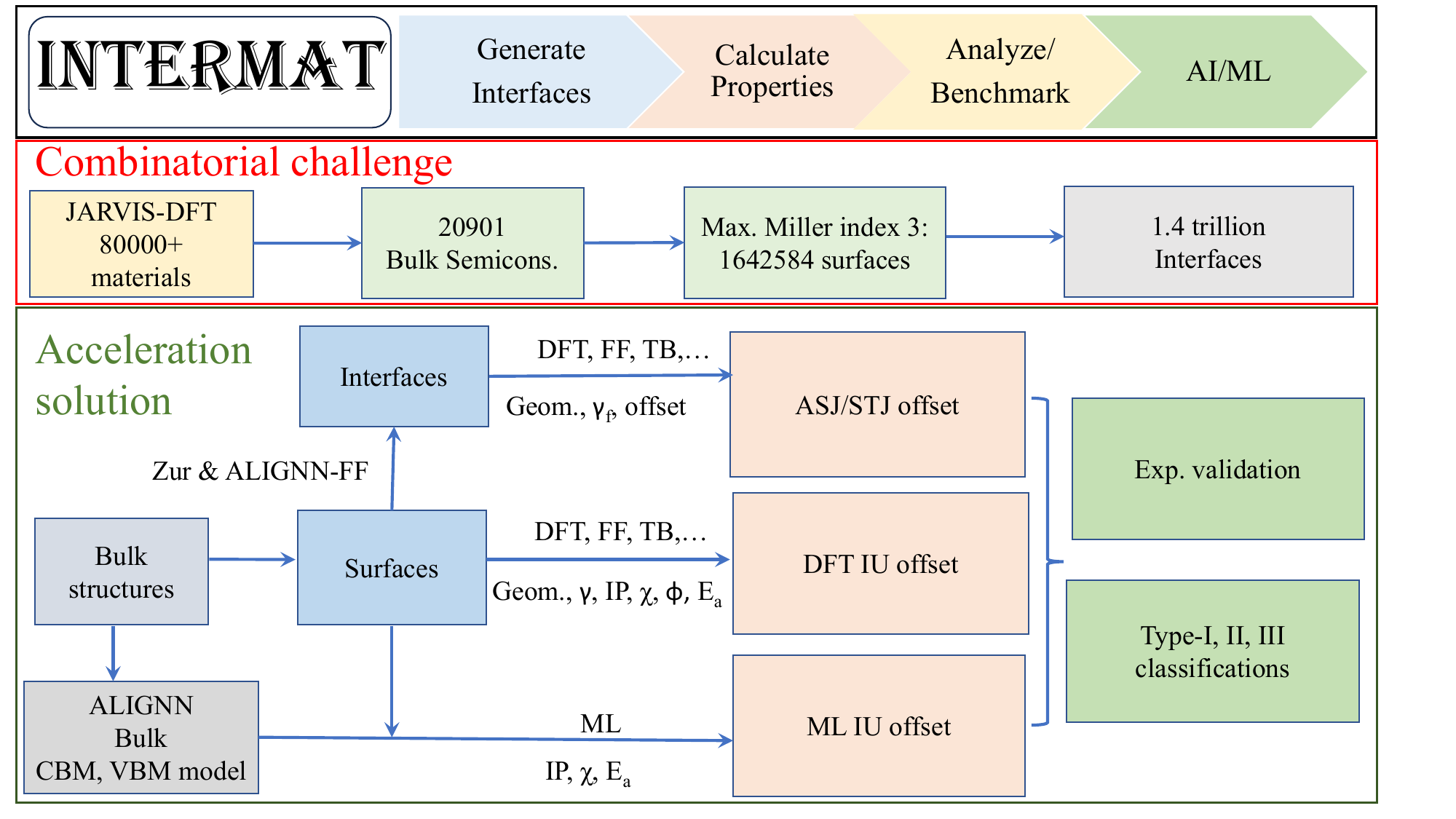}
\caption{\textcolor{black}{Schematic overview of the workflow. InterMat can be used to generate surface and interface geometric structures (Geom.), perform multi-fidelity calculations (such as density functional theory, force-field, tight-binding and machine learning) to predict properties (such as surface energy, interface formation energy, band offset, work function, ionization potential, electron affinity), analyze and benchmark data against experiments, and utilize machine learning models for such data. The number of possible semiconductor-semiconductor interfaces is exceedingly large. The workflow aims to provide a toolkit to generate interface structures and use multi-fidelity methods to accelerate interface/heterostructure design.}}
\label{fig:schematic}
\end{figure}

\begin{figure}[hbt!]
\centering
\includegraphics[width=\linewidth]{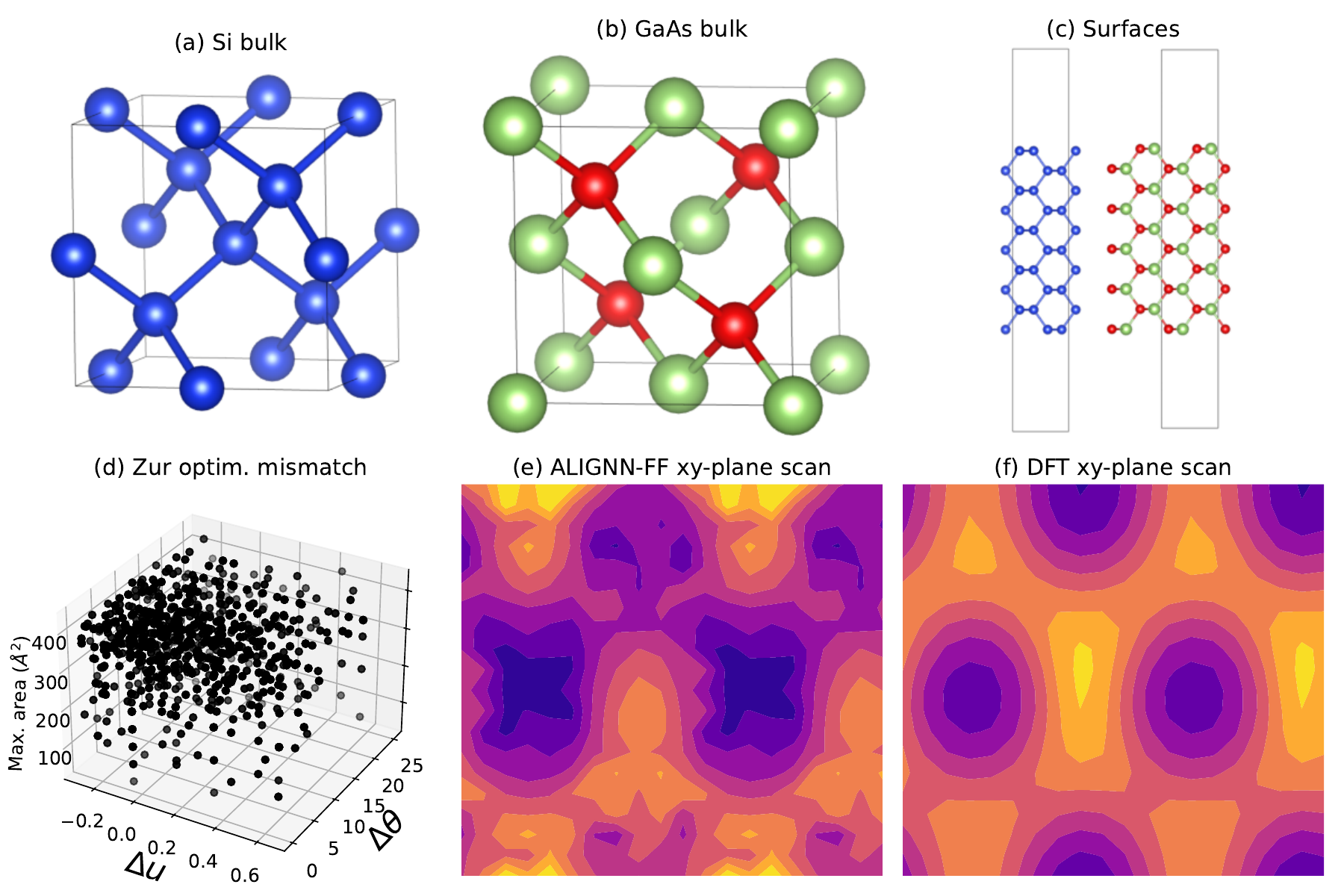}
\caption{\textcolor{black}{Structure generation criteria selection and initial xy-plane scan with ALIGNN-FF for Si(110)/GaAs(110) interface. a) atomic structure of silicon (Si), b) atomic structure of gallium arsenide (GaAs), c) surfaces (110) generated from the bulk structures of Si (left) and GaAs (right), d) candidate interface parameters from Zur algorithm: mismatch in x-direction (u), mismatch in y-direction (v) and maximum allowed area to generate suitable structures. e) ALIGNN-FF and f) DFT energy as a function of displacement in xy-plane of interface.}}
\label{fig:zsl}
\end{figure}

\begin{figure}[hbt!]
\centering
\includegraphics[width=\linewidth]{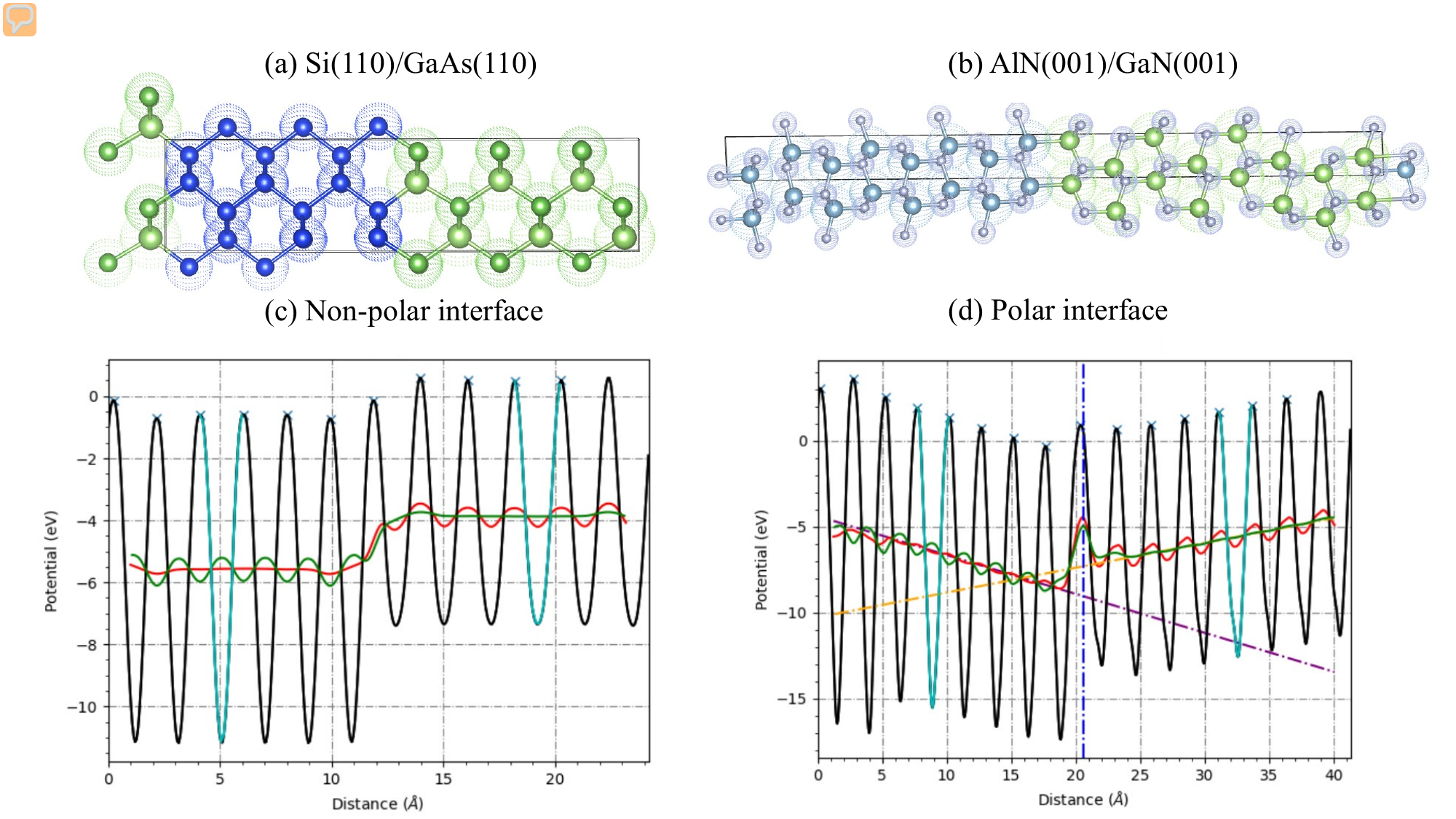}
\caption{Atomic structures and band-alignment using the average electrostatic potential of semiconductor interfaces. a) atomic structure view of Si/GaAs(110), b) atomic structure view of polar interface AlN/GaN(001), c) electrostatic potential profile for non-polar interface Si/GaAs(110), d) average electrostatic potential profile for polar interface of AlN/GaN (001). The cyan lines are used to find the repeat unit layers for the left and right parts. The red and green lines show the potential averaged over the repeat distances of the left and right slabs, respectively. The dotted vertical blue line marks the interface.}
\label{fig:band_alignn}
\end{figure}

\begin{figure}[hbt!]
\centering
\includegraphics[width=\linewidth]{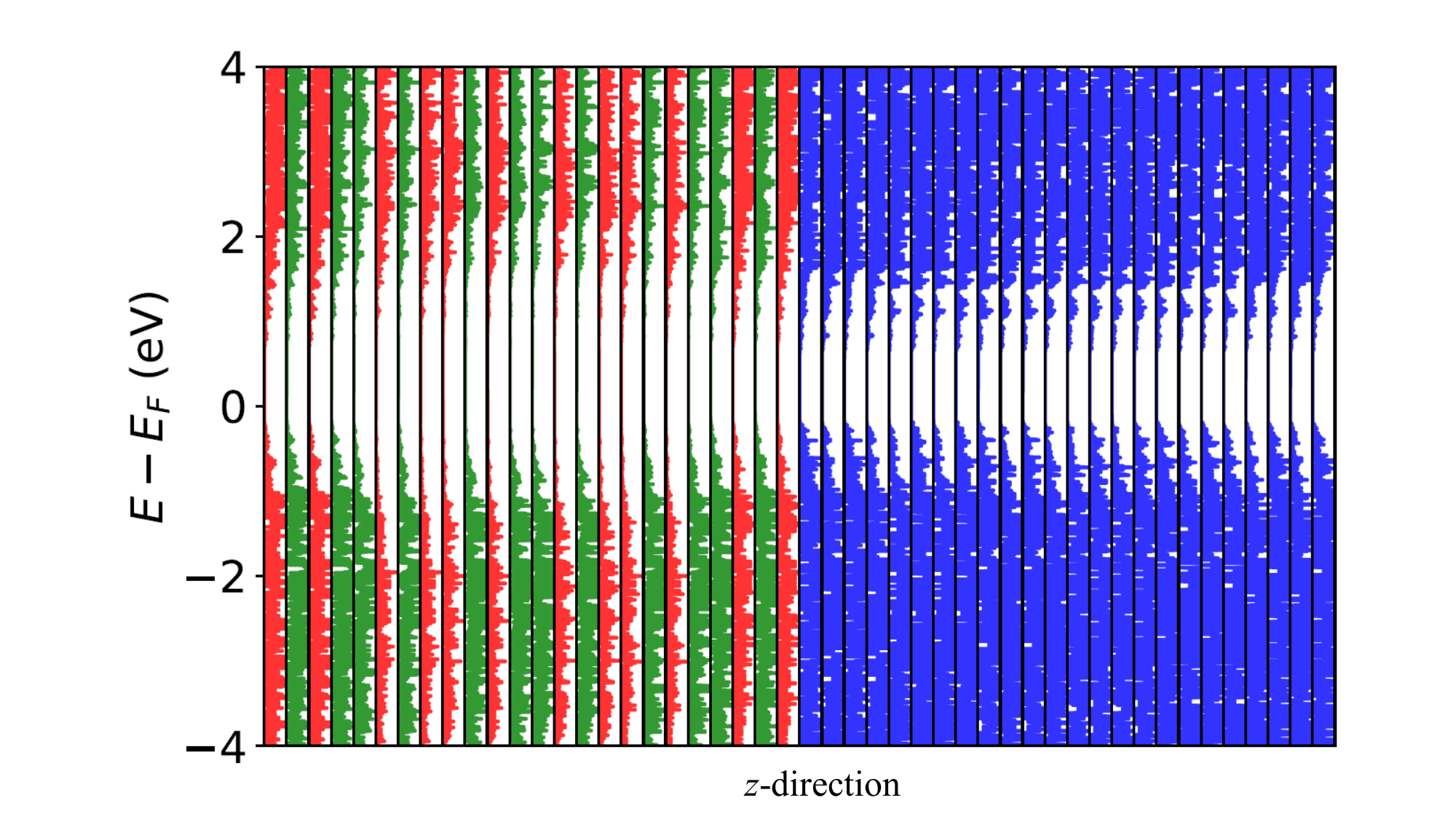}
\caption{Atom projected density of states for the system, separated along the z direction, normal to the interface. Red, green and blue colors represent gallium, arsenic and silicon atom contributions respectively.}
\label{fig:atdos}
\end{figure}

\begin{figure}[hbt!]
\centering
\includegraphics[width=\linewidth]{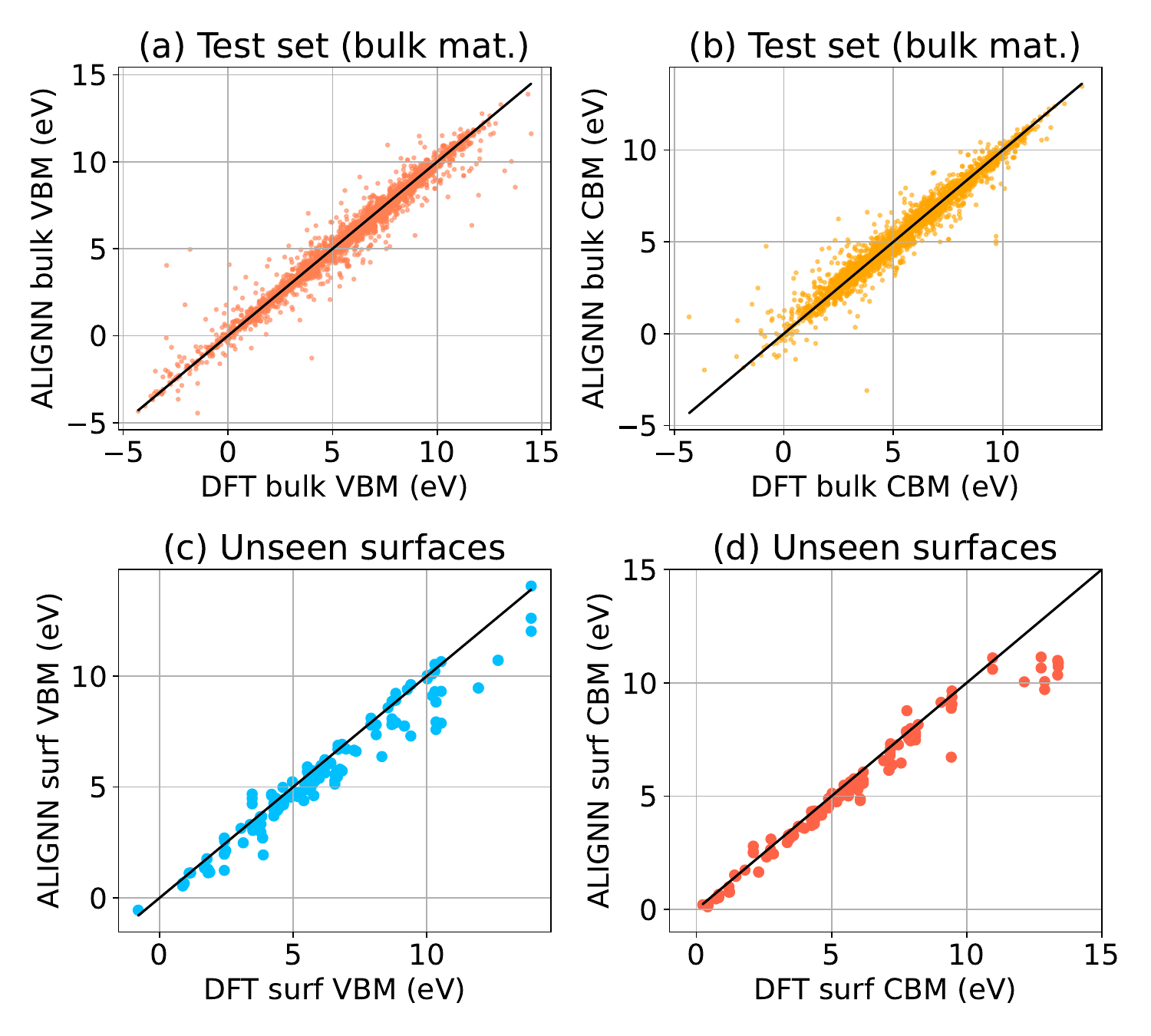}
\caption{ALIGNN based regression models for a) VBM for JARVIS-DFT 3D/bulk materials test set data, b) CBM for bulk materials test set data, c) slab surface VBMs not part of the training set to evaluate extrapolation strength, d) slab surface CBMs not part of the training set to evaluate extrapolation strength.}
\label{fig:deeplearn}
\end{figure}

\begin{figure}[hbt!]
\centering
\includegraphics[width=\linewidth]{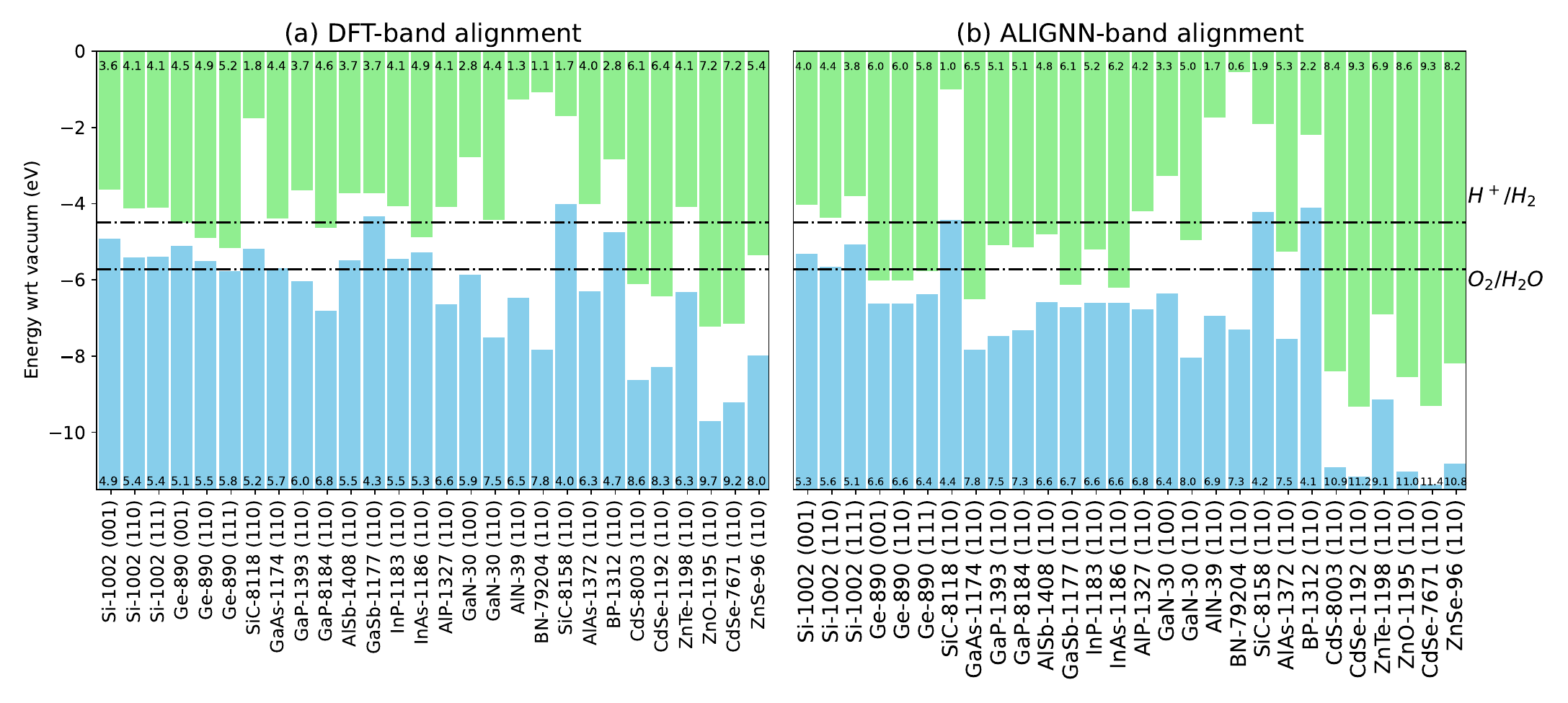}
\caption{A few examples of band alignment based on the ionization potential (IP) and electron affinity (EA) for Anderson's/independent unit (IU) model. a) Density functional theory based alignments can be used to obtain band offsets. b) Fast ALIGNN based band alignment predictions. The numbers in the green bars represent electron affinity while that in the blue bars represent ionization potentials. The trained models based band alignment will be available at JARVIS-Heterostructure website soon. }
\label{fig:offsetexample}
\end{figure}

\begin{figure}[hbt!]
\centering
\includegraphics[width=\linewidth]{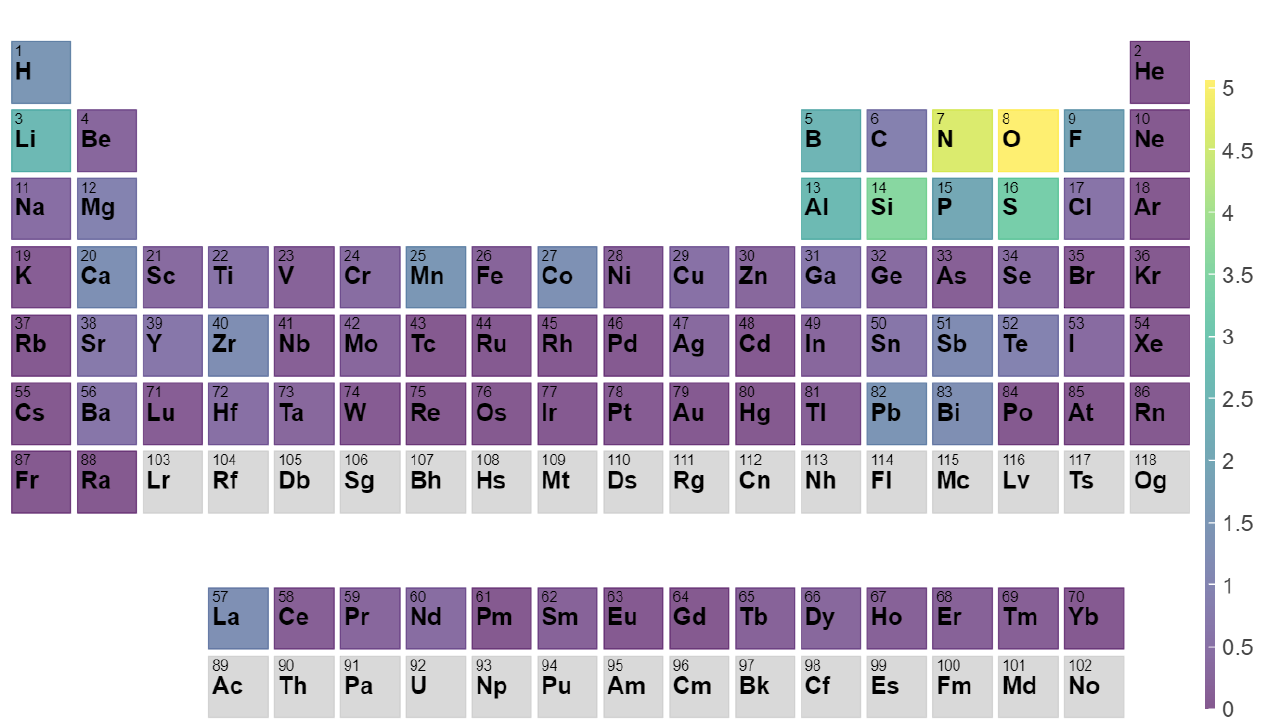}
\caption{
Percentage chance that a heterostructure with a given element will form a type-I heterostructure with silicon as the second semiconductor.}
\label{fig:ptable}
\end{figure}

A schematic overview of InterMat along with the combinatorial problem of interfaces is shown in Fig. \ref{fig:schematic}. \textcolor{black}{InterMat can be used to generate surface and interface structures, perform multi-fidelity calculations to predict properties, analyze and benchmark data against experiments, and train and utilize machine learning models based on the resulting data.} Initial atomic structures can be obtained from the JARVIS-DFT repository. As an example of the combinatorial challenge of interfaces, we can consider starting with the 20901 semiconductors in the JARVIS-DFT databsae with OptB88vdW band gaps between 0.1 eV and 6 eV. Using a maximum Miller index (M) of 1, 2, and 3, the number of symmetrically distinct surface slabs are 186847, 591639 and 1642584, respectively. Using these surfaces, the number of binary interface systems that can be generated are 17.5 billion, 175 billion and 1.4 trillion. Including the possibility of different atomic surface terminations, reconstructions, and defects further complicates matters. 

It is unrealistic to analyze such a large search space to find combinations for device applications by using conventional experimental or computational techniques. We will instead use ALIGNN models trained \textcolor{black}{on bulk materials} to guide and prioritize DFT calculations among the vast pool of candidate surfaces and interfaces. We develop machine learning models for fast predictions of valence band maxima (VBM) and conduction band minima (CBM) using ALIGNN \textcolor{black}{on JARVIS-DFT bulk material dataset}. These predictions can be further used for fast IU based band alignment using electron affinity/Anderson's rule \cite{anderson1960germanium}. To assess the strengths and limitations of such models, we develop a surface dataset for independent unit (IU) models and an interface dataset for alternate slab junction (ASJ)/superlattice models using DFT. We particularly focus on industrially relevant semiconductors including group IV (C, Si, Ge etc.), III-IV (AlN, GaN, GaAs, GaP, InSb etc.), II-VI (CdS, CdSe, ZnO, ZnS etc.). We also assess the strengths and limitations of IU and ASJ models against experimental measurements. This DFT dataset can then be fed back into the ALIGNN models to further improve accuracy. 

\subsection*{DFT surface dataset: work function, electron affinity, ionization potential and surface energy}



\begin{table}[hbt!]
\centering
\caption{Work function ($\phi$, eV), electron affinity ($\chi$, eV) and surface energy ($\gamma$, $Jm^{-2}$) of a few unreconstructed non-polar surface slabs from OptB88vdW (OPT) against experimental data. The IDs represnt JARVIS-DFT identifiers.}
\label{table:wf}
\begin{tabular}{||c c c c c c c c c||} 
 \hline
 System & IDs & Miller & $\phi$ (OPT)&$\phi$ (Exp) & $\chi (OPT)$ &$\chi $(Exp) & $\gamma$ (OPT) & $\gamma$ (Exp) \\  
 \hline
Si & 1002& 111  & 5.00&4.77\cite{dillon1958work} &4.10 &4.05\cite{bhattacharya1997semiconductor} &1.60 &1.14\cite{messmer1981surface} \\
Si & 1002& 110  & 5.30&4.89\cite{dillon1958work} &4.10 & -&1.66 &1.9\cite{messmer1981surface}\\
Si & 1002& 001  & 5.64&4.92\cite{dillon1958work} &3.60 & -&2.22 &2.13\cite{jaccodine1963surface}\\
C & 91& 111  & 4.67&5.0\cite{holzl2006work} & -2.9& -&5.27 &5.50\cite{field1981strength}\\
Ge & 890& 111  & 4.87&4.80\cite{gobeli1964photoelectric} &5.2 & 4.13\cite{milnes2012heterojunctions}& 0.99&1.30\cite{jaccodine1963surface}\\
SiGe & 105410& 111  & 4.93&4.08\cite{pouch2015work} &4.5 &- & 1.36&-\\
SiC & 8118& 001  & 5.26&4.85\cite{pelletier1984application} &1.3  &- &3.51 &-\\
GaAs & 1174& 110  & 4.89&4.71\cite{liu2007first} & 4.40 &4.07\cite{bhattacharya1997semiconductor} & 0.67&0.86\cite{messmer1981surface}\\
InAs & 1186& 110  & 4.85&4.90\cite{liu2007first} & 4.9 & 4.9\cite{milnes2012heterojunctions}& 0.57&-\\
AlSb & 1408& 110  & 5.11&4.86\cite{liu2007first} &  3.70&3.65\cite{milnes2012heterojunctions} & 0.77&-\\
GaSb & 1177& 110  & 4.48&4.76\cite{liu2007first} & 3.70 &4.06\cite{milnes2012heterojunctions} &0.71 &-\\
AlN & 39& 100 & 5.56&5.35\cite{pelletier1984application} & 1.3 & 2.1 \cite{wu1998electron}& 2.27&-\\
GaN & 30& 100   & 5.74&5.90\cite{rosa2006first} & 2.8  &3.3\cite{lin2012experimental}& 1.67&-\\
BN & 79204& 110   & 6.84&7.0\cite{lu2022towards} &  1.4 &- & 2.41&-\\
GaP & 1393& 110   & 5.31&6.0\cite{pelletier1984application} &  4.0&  4.3\cite{bhattacharya1997semiconductor}& 0.88&1.9\cite{messmer1981surface}\\
BP & 1312& 110   & 5.61&5.05\cite{crovetto2022boron} & 2.8 & - & 2.08&-\\
InP & 1183& 110  & 5.17&4.65\cite{liu2007first} &  4.10&4.35\cite{bhattacharya1997semiconductor} &0.73 &-\\
CdSe & 1192 & 110  & 5.70&5.35\cite{csik2005density} &6.4  &- & 0.38&-\\
ZnSe & 96 & 110  & 5.67&6.00\cite{haase1990characterization} &5.4  &- &0.44 &-\\
ZnTe & 1198 & 110  & 5.17&5.30\cite{shen2020insights} &4.10  &3.5\cite{milnes2012heterojunctions} & 0.36&-\\
Al & 816& 111  & 4.36&4.26\cite{holzl2006work} &- &- & 0.82&-\\
Au & 825& 111  & 5.5&5.31\cite{holzl2006work} &  -&- & 0.90&-\\
Ni & 943& 111  & 5.35&5.34\cite{holzl2006work} &- & -& 2.02&2.34\cite{clark1980effect}\\
Ag & 813& 001  & 4.5&4.2\cite{holzl2006work} & -&- & 0.99&-\\
Cu & 867& 001 & 4.7&5.1\cite{holzl2006work} & -&- & 1.47&-\\
Pd & 963& 111  & 5.54&5.6\cite{holzl2006work} & -&- & 1.57&-\\
Pt & 972& 001  & 5.97&5.93\cite{holzl2006work} & -&- & 1.94&-\\
Ti & 1029& 100  & 3.84&4.33\cite{holzl2006work} & -&- & 2.27&-\\
Mg & 919& 100 & 3.76&3.66\cite{holzl2006work} & -& -& 0.35&-\\
Na & 931& 001  & 2.97&2.36\cite{holzl2006work} & -&- & 0.10&-\\
Hf & 802& 111  & 3.7&3.9\cite{holzl2006work} & -&- & 2.02&-\\
Co & 858& 001  & 5.22&5.0\cite{holzl2006work} & -&- & 3.49&-\\
Rh & 984& 001  & 5.4&4.98\cite{holzl2006work} & -&- & 2.46&-\\
Ir & 901& 100  & 5.85&5.67\cite{holzl2006work} & -&- & 2.77&-\\
Nb & 934& 100  & 3.87&4.02\cite{holzl2006work} & -&- & 2.41&-\\
Re & 981& 100  & 4.96&4.72\cite{holzl2006work} & -&- & 2.87&-\\
Mo & 21195& 100 & 4.17&4.53\cite{holzl2006work} & -&- & 3.30&\\
Zn & 1056& 001  & 4.27&4.24\cite{ashcroft2022solid} & -&- & 0.36&-\\
Bi & 837& 001  & 4.31&4.34\cite{holzl2006work} & -&- & 0.65&0.43\cite{tran2016surface}\\
Cr & 861& 110  & 5.04&4.5\cite{holzl2006work} & -&- & 3.31&-\\
Sb & 993& 001  & 4.64&4.7\cite{holzl2006work} & -&- & 0.67&-\\
Sn & 1008& 110 & 4.82&4.42\cite{holzl2006work} & -&- & 0.91&-\\
\hline
MAE & -& - & 0.29&-& 0.39&- & 0.34&-\\


\hline
\end{tabular}
\end{table}

We develop a dataset of non-polar unreconstructed slab surfaces using the JARVIS-DFT workflow and bulk material dataset. Examples of silicon and gallium arsenide bulk atomic structures are shown in Fig. \ref{fig:zsl}a and Fig. \ref{fig:zsl}b respectively. Next, we generate surface slab structures with a thickness of 1.6 nm and vacuum padding of 1.2 nm, as shown in Fig. \ref{fig:zsl}c. Recently, it was shown that vacuum and slab thicknesses of at least 10 Å are sufficient for surface models \cite{tran2016surface}. During the DFT calculations, the converged $k$-point \cite{choudhary2019convergence} values from the relevant bulk calculations are used for surfaces. We optimize the internal coordinates of these surfaces keeping the cell volume constant. 

We carefully benchmark surface energy ($\gamma$), ionization potential (IP), electron affinity ($\chi$), and work function ($\phi$) values against experimental measurements from the literature. The surface energy ($\gamma$) can be calculated using the formula:

\begin{equation} 
    \gamma = \frac{E_{\text{slab}} - N_{\text{bulk}} \cdot E_{\text{bulk}}}{2A} 
\end{equation} 
\textcolor{black}{where $(E_{\text{slab}})$ is the total energy of the relaxed slab model, $(N_{\text{bulk}})$ is the number of bulk-like atoms in the slab model, $(E_{\text{bulk}})$ is the energy per atom in the bulk material, and $(A)$ is the surface area of the slab model. The factor of 2 accounts for the fact that there are two surfaces in the non-polar slab model (top and bottom).}

We obtain the valence band maximum (VBM) and vacuum level ($E_{vac}$) of surface slabs from DFT calculations using the OptB88vdW functional. \textcolor{black}{The work function is obtained by subtracting the vacuum level from the Fermi level ($\phi=E_{vac}-E_F$)}. Similarly, the ionization potential is the difference between the VBM and $E_{vac}$. Then, we add the electronic bandgap ($E_g$) of the \textcolor{black}{bulk} material to the ionization potential to get the electron affinity (EA, $\chi$). \textcolor{black}{Semi-local DFT has proven quite effective in describing the valence bands of materials but is known to underestimate band gaps. For accurate prediction of both valence and conduction bands, particularly in materials with complex electronic interactions, higher-level theories like many-body perturbation theory (\textit{e.g.} GW calculations) might be necessary, but are computationally very expensive. In order to address this problem in a more computationally efficient manner, we make use of bulk band gaps from the JARVIS-DFT database computed using the TBmBJ metaGGA functional. TBmBJ predictions can provide band gap descriptions with accuracy close to more expensive methods but at an order of magnitude less computational cost\cite{choudhary2018computational}, which is important for high-throughput studies. We calculate surface conduction band quantities by first calculating the valence band at the GGA-level using OptB88vdW and then add to that the TBmBJ bulk gap to get the conduction band minimum (CBM). Performing full surface calculations using hybrid functionals or GW is beyond the scope of the present work because of excessive computational cost, but we plan to provide further tests of those approaches in the future \cite{ghosh2022efficient,hinuma2014band}.} 

In order to benchmark our surface dataset, we compare work functions, electron affinities, and surface energies of several dozen surfaces with experimental data in Table. \ref{table:wf}. We find excellent agreement for the work functions, with a mean absolute error value of 0.29 eV, consistent with previous benchmarking efforts \cite{de2016error}. Similarly, we obtain a mean absolute error of 0.39 eV and 0.34 $Jm^{-2}$ for the electron affinity and surface energy, respectively. Currently, we have performed calculations on \nsurfaces surfaces using the workflow, and the dataset is still growing. Using, these \nsurfaces surfaces, \iuband IU-band offsets can be predicted. Also, we plan to include reconstructed and polar surfaces in the future.

\subsection*{DFT Interface dataset: alternate slab junction (ASJ) band alignment}



We next consider explicit DFT calculations of interfaces, which first require generating candidate interface structures. This can be done in either of the following two ways: 1) by attaching the two surface slabs together without vacuum padding, creating a superlattice or alternating slab junction (ASJ) structure, or 2) by attaching the two surface slabs with vacuum padding, creating a surface terminated junction (SJT) structure. We have focused on the ASJ approach\cite{di2021band, nano11061581}. \textcolor{black}{After obtaining surface slab structures as discussed in the previous section,} we generate the interfaces following the Zur \textit{et. al.} algorithm \cite{zur1984lattice}. The Zur algorithm generates a number of superlattice transformations within a specified maximum surface area and also evaluates the length and angle between film and substrate superlattice vectors to determine if they can match within a tolerance. This algorithm is applicable to different crystal structures and their surface orientations. We use a maximum lattice mismatch of 8 \%, maximum area of 300 ${\textup{\AA}}^2$, and maximum angle tolerance of 1 degree. Note that in previous studies, lattice mismatch of 20 \% has been reported\cite{hinuma2014band,goodhew1999strain}. An example of the application of the algorithm to the Si(110)/GaAs(110) interface is shown in Fig. \ref{fig:zsl}d with several lattice length and angle mismatches ($\Delta u$ and $\Delta \theta$) as well as maximum area. After eliminating structures with area higher than max-area tolerance and structures with mismatch angle more than the specified angle threshold, we then choose the remaining structure (if any) with the minimum mismatch lattice vector lengths.  

The Zur algorithm determines a candidate unit cell, but the relative alignment of the structures in the in-plane, as well as the slab terminations still need to be decided. For the in-plane alignment, we perform a grid search of possible options with a spacing interval of 0.05 fractional coordinates to determine the initial structure for further relaxation. Doing such a large number of calculations with DFT would be prohibitive, so we use ALIGNN-FF \cite{choudhary2023unified} to identify the starting in-plane alignment. ALIGNN-FF is a universal force field ML model developed using JARVIS-DFT data with 307113 structures and can be used to model combinations of 89 elements from the periodic table. An example of ALIGNN-FF predictions of an in-plane grid search is shown in Fig. \ref{fig:zsl}e. \textcolor{black}{For the Si/GaAs(110) case, we also perform corresponding DFT calculations as shown in Fig. \ref{fig:zsl}f. Here high-peaks (yellow color using magma colormap) usually represent too close atoms during the translation operations, which should be avoided. Clearly, the DFT contours are smoother than ALIGNN-FF because of its relatively rough potential energy surface (PES). Nevertheless, the minimums of the contours, which are of interest for in-plane alignments, closely resemble each other. As the ALIGNN-FF accuracy increases with more data, we expect to get much smoother PES in future.} After the ALIGNN-FF calculations to select the initial alignment, a full DFT relaxation is performed.

For computational purposes, it is important to have a unique identifier for an interface. While generating the interfaces, we use a naming convention to include a) material IDs (such as JVASP-1002 for Si and JVASP-1174 for GaAs), b) film and substrate Miller indices (such as 110 for each), c) film and substrate thickness values (such as 16 ${\textup{\AA}}$ each), d) separation between these two surface slabs (such as 2.5 ${\textup{\AA}}$ for an ASJ model, 18 ${\textup{\AA}}$ for STJ interface models), e) relative displacement in xy-plane (such as a displacement vector of [0.5, 0.2]), f) calculator method (such as DFT (VASP), ALIGNN-FF etc.) giving rise to an interface with a name such as: Interface-JID1\_JID2\_film\_miller\_M1\_sub\_miller\_M2\_film\_thickness\_T1\_subs\_thickness\_T2\_separation\_S\_disp \_X\_Y\_vasp (where JID1 is JVASP-1002, JID2 is JVASP-1174, M1 and M2 are both 1\_1\_0, T1 and T2 are 16, S is 2.5, X is 0.5, Y is 0.2). Such a scheme helps to reproduce the unique interfaces. Of course, realistically, more complex parameters for an interface can be important such as terminations, reconstructions, misfit-dislocation, vacancies on the interface etc., but they can be easily included in the naming scheme as well later.

After selecting a good guess of the interface using the above approach, DFT calculations are performed to calculate \textcolor{black}{quantities such as the interface formation energy and }the band offset value. During the DFT calculations, the more converged $k$-point grids and energy cutoffs of the two constituent bulk materials\cite{choudhary2019convergence} is used. An example of the Si(110) and GaAs(110) interface is shown in Fig. \ref{fig:band_alignn}a. Furthermore, we can project the electron density of states across the cell dimension in Fig.\ref{fig:atdos} to show how electronic states are distributed along the interface region. We observe the GaAs gap decrease near the silicon region. Such analysis can help to understand the local band alignment and atomic character, which are important for device modeling.

After determining the optimized geometric structure for the interface using DFT, we obtain  \textcolor{black}{ the interface formation energy and valence} band offset data using the formalism detailed in Ref. \cite{romanyuk2016ab} and Ref. \cite{van1985theoretical} respectively. As an example, we show a detailed analysis of Si(110)/GaAs(110) and AlN(001)/GaN(001) in Fig. \ref{fig:band_alignn}. \textcolor{black}{The interface formation energy ($\gamma_f$) is calculated using the formula:}
\begin{equation} 
  \gamma_f A= E_{tot}-  \sum_i n_i \mu_i
\end{equation} 
\textcolor{black}{
where $\gamma$ interface formation energy, $E_{tot}$ is the total energy of the superlattice, $\mu_i$ is the chemical potential of the specie $i$, $n_i$ is the number of atoms of the specie $i$, and $A$ is the interface unit cell area. Using the bulk materials energy per atom in its most stable form in JARVIS-DFT and OptB88vdW functional, we obtain an interface formation energy of -0.056 J$m^{-2}$ for the Si(110)/GaAs(110) system. A negative formation energy suggests a feasible formation of the interface. Moreover, such interface formation energies with varying chemical potentials of the constituent elements can provide information about the thermodynamic stability of the interface in different growth conditions. Such detailed tasks for individual interfaces will be carried in future.}

In Fig. \ref{fig:band_alignn}a, we show the atomic structure of the ASJ based heterostructure of Si(110)/GaAs(110). The left side (with blue atoms) represents the Si and the right side is the GaAs region. In Fig. \ref{fig:band_alignn}c, we show the electrostatic potential profile, averaged in-plane, of the interface. The approximately sinusoidal profile on both regions represents the presence of atomic layers. The cyan lines show the region used to define the repeat distance, $L$, used for averaging in each material (see below). The red and green lines show the average potential profiles for the left and right parts using the repeat distance. The valence band offset ($\Delta E_v$) of an interface between semiconductor A and B, $\Delta E_v$ is obtained using eq. 4. The difference in the averages for the left and right parts gives the $\Delta V$ term. Now the bulk VBMs of the left and right parts are also calculated to determine the $\Delta E$. The sum of these two quantities gives the valence band offset that can be compared to experiments. 

\begin{equation} 
 \Delta E_v (A/B)= (E_v^B-E_v^A) + \Delta V
\end{equation} 

\begin{equation} 
  \Delta V = \bar{\bar{V}}_A - \bar{\bar{V}}_B
\end{equation} 
Here, ${E_v^A}$ (${E_v^B}$) represents the position of the VBM with respect to the average electrostatic potential in the bulk material A (B), and $\Delta V$ represents dipole potential or the difference between the macroscopic-averaged electrostatic potential between A and B. 
Moreover, $\bar{\bar{V}}$ is the average along the repeat unit $L$ of $\bar{V}$, which is the planar averaged electrostatic potential:

\begin{equation} 
  \bar{\bar{V}}(z) = \frac{1}{L}  \int_{-L/2}^{L/2} \bar{V}(z+z^{'}) dz^{'} 
\end{equation} 
$\bar{V}$ is given by:
\begin{equation} 
  \bar{V}(z) = \frac{1}{S}  \int_{S} V(x,y,z)dxdy 
\end{equation} 
where, $L$ is the distance between repeat units and $S$ is the area which is parallel to the interface. 
The corresponding conduction-band offset can be determined by using TBmBJ band-gap values from the respective bulk calculations or experimental data. We will use the convention that a positive value of the valence-band offset at an interface A/B indicates that the VBM is higher in material B. For the GaAs/Si interface we obtain $\Delta E_v$ of 0.31 eV and 0.39 eV using OptB88vdW and R2SCAN functionals, respectively, which is in close agreement with the experimental value of 0.23 eV.

\begin{table}[hbt!]
\centering
\caption{Valence band offsets (in eV) of a few independent unit (IU)/Anderson’s model and alternating slab-junction  (ASJ) based semiconductor/semiconductor interfaces with OptB88vdW (OPT) and R2SCAN functionals in comparison to previously reported experiments. Here ID, Miller and P represent a JARVIS-DFT identifier, Miller index and polar surface interfaces respectively.}
\label{table:example}
\begin{tabular}{||c c c c c c c ||} 
 \hline
 System & ID & Miller& IU (OPT) &ASJ (OPT) & ASJ (R2SCAN)&Exp  \\  
 \hline

AlP/Si & 1327/1002& 110/110 & 1.24& 0.88 &1.04&1.35 \cite{di2021band}  \\ 
GaAs/Si & 1174/1002 &110/110 & 0.30& 0.31 &0.39&0.23 \cite{list1987si}\\ 
CdS/Si& 8003/1002 &110/110 & 3.22& 1.48 &1.70&1.6 \cite{kundu1993chemical} \\

AlAs/GaAs & 1372/1174 &110/110 & 0.60& 0.48&0.50&0.55 \cite{batey1986energy}\\
CdS/CdSe & 8003/1192 &110/110 & 0.35& 0.10&0.11& 0.55 \cite{talapin2003highly}\\
InP/GaAs & 1183/1174 &110/110 & 0.25& 0.72 &0.75& 0.19\cite{waldrop1989measurement}\\
ZnTe/AlSb & 1198/1408 &110/110 & 0.8& 0.25&0.33& 0.35 \cite{schwartz1990band}\\
CdSe/ZnTe & 1192/1198& 110/110 &1.8 & 0.58&0.67&0.64\cite{yu1991measurement} \\
InAs/AlAs & 1186/1372 &110/110 &- & 0.46&0.39&0.5 \cite{arriaga1991electronic}\\
InAs/AlSb & 1186/1408 &110/110 & -& 0.05&0.16& 0.09 \cite{nakagawa1989electrical}\\
ZnSe/InP & 96/1183 &110/110 &- & 0.13 &0.18&0.41 \cite{lange2020spectroscopic}\\ 
InAs/InP & 1186/1183 &110/110 &- & 0.11 &0.09& 0.31\cite{waldrop1989measurement}\\
ZnSe/AlAs & 96/1372 &110/110 &- & 0.38 &0.45& 0.4 \cite{rubini1998transitivity}\\
GaAs/ZnSe & 1174/96 &110/110 &-& 0.72 &0.80 & 0.98 \cite{kowalczyk1982measurement}\\
ZnS/Si& 10591/1002 &001/001 &-& 0.92 & 1.16&1.52 \cite{lew1997electronic} \\
Si/SiC & 1002/8118& 001/001 &-& 0.51 &0.47&0.5 \cite{dufour1997sic}\\
GaN/SiC (P) & 30/8118& 001/001 &-&  1.12&1.37&0.70 \cite{rizzi1999aln}\\ 
Si/AlN (P) & 1002/30& 001/001 &-& 3.51 &3.60&3.5 \cite{king2015band}\\
GaN/AlN (P) & 30/39& 001/001 &-&  0.80&0.86&0.73 \cite{sang2014band} \\
AlN/InN (P) & 39/1180& 001/001 &-&  1.24&1.07&1.81 \cite{waldrop1996measurement} \\
GaN/ZnO (P) & 30/1195& 001/001 &-& 0.51 &0.46&0.7 \cite{liu2011band} \\
\hline

MAE & -& - &0.45& 0.22 &0.23&- \\

\hline
\end{tabular}
\end{table}

Next, we show a polar semiconductor heterojunction example for AlN (001)/GAN(001) interface in Fig. \ref{fig:band_alignn}b. The electrostatic potential profile is shown in Fig. \ref{fig:band_alignn}d. In contrast to flat average potential values in Fig. \ref{fig:band_alignn}c, we observe inclined profiles for this system indicating the presence of a constant electric field. We fit lines for both sides and extrapolate to the interface. The difference of the lines at the interface gives $\Delta V$. The calculation of $\Delta E$ remains the same as the non-polar case. These calculations are automated in the workflow, however, it is important to check that the slab is thick enough to define a bulk-like region where $\bar{\bar{V}}(z)$ is linear. Now, in the Table. \ref{table:example} we compare some of the ASJ based valence band offsets ($\Delta E_v$) with experimental measurements. We find a mean absolute error of 0.22 eV and 0.23 eV for OptB88vdW and R2SCAN respectively, which is comparable a value of 0.16 eV from to Liberto \textit{et. al.} for a smaller number of systems using the HSE06 functional. In the future, we plan to carry out HSE06 calculations for surfaces as well as interfaces to further improve the quality of predictions. These benchmarks will also be available in the JARVIS-Leaderboard platform \cite{choudhary2023large} as well. Out of numerous possible combinations, only \nasj DFT calculations of ASJ-based interfaces are available right now and the database is still growing.

\subsection*{DFT-based independent unit (IU) band alignment}

IU band alignment, also known as Anderson's rule \cite{anderson1960germanium}, predicts semiconductor band offsets at interfaces using only the IP and EA data from independent surface calculations. For a semiconductor heterojunction between A and B, the conduction band offset is given by: 
\begin{equation} 
  \Delta E_c = \chi_B -  \chi_A
\end{equation} 
Similarly, the valence band offset is given by:
\begin{equation} 
  \Delta E_v = (\chi_A+ E_{gA}) -  (\chi_B+ E_{gB})
\end{equation} 
In Fig.\ref{fig:offsetexample}a, we show the DFT-based IU band alignments for a set of well-known semiconductor surfaces.
We also include dotted lines for the energy levels of H$_2$ and H$_2$O, which are relevant for photo-catalyst applications. We compare the DFT based IU band offsets for 8 interfaces in Table \ref{table:example} against experiments. We find a mean absolute error of 0.45 eV which is similar to a value of 0.32 eV as found in ref. \cite{di2021band} for different systems.


\subsection*{ALIGNN-based IU alignment from bulk data}

We seek to accelerate the prediction of band edges using ML models, but the absolute prediction of band edges relative to vacuum requires DFT calculations with surfaces, which are too computationally expensive to create a robust dataset. JARVIS-DFT contains a much larger dataset of three dimensional materials with CBM and VBM values. \textcolor{black}{Here, the CBMs and VBMs are simply the band edges written out by VASP for the bulk materials dataset using OptB88vdW. These band edges use the VASP convention that the average electrostatic potential of a unit cell is set to zero, and are not directly comparable to experimental values. A surface calculation with explicit vacuum is necessary to align the VBM/CBM to vacuum. We first train an ML model using ALIGNN based on these bulk quantities, but we will then show that this model is somewhat surprisingly also useful for predictions of band edges relative to vacuum. }



To train the ALIGNN model, we split each bulk VBM/CBM dataset into 90:5:5 train:validation:test parts. We train on 90 \% train data and evaluate the validation and test data using ALIGNN. We find a mean absolute error (MAE) of 0.28 ev for CBM and VBM. We note that the CBM/VBM data can vary from -10 to 10 eV (as shown in Fig. \ref{fig:deeplearn}a,b) suggesting that the model should be reasonable for predictions. The mean absolute deviation (MAD) for the CBM and VBM are 2.08 eV and 2.67 eV, respectively, so the MAD:MAE is nearly 10 relative to a trivial baseline model. Out of several other material properties trained using ALIGNN \cite{choudhary2021atomistic}, CBM/VBM models has one of the highest MAD:MAE ratios, especially compared to other electronic properties like the band gap. We show the CBM and VBM prediction models in Fig.\ref{fig:deeplearn}a and Fig.\ref{fig:deeplearn}b respectively. We believe with more data using the active learning loop we can further increase the MAD:MAE in future \cite{choudhary2022recent}.

Next, we evaluate the bulk-trained ALIGNN models on the DFT surface dataset and show results in  Fig.\ref{fig:deeplearn}c and  Fig.\ref{fig:deeplearn}c for CBM and VBM respectively. We note that the training data does not include any surfaces. Nevertheless, most of the data points are on x = y line suggesting excellent agreement. We find MAE values of 0.55 eV and 0.96 eV for VBM and CBM respectively. \textcolor{black}{This level of agreement is surprising because the value of the averaged electrostatic potential will change as the ratio of vacuum thickness to slab thickness changes. We also note that the error in CBM is higher than that of VBM, perhaps an explanation for why predicting band gaps of materials using machine learning is ever-standing difficult problem. }


\textcolor{black}{Up to this point, our model can only predict quantities relative to a cell-averaged electrostatic potential, which cannot be directly compared to experiment. However, we observe that our ALIGNN model predictions of the bulk VBM/CBM are in fact strongly correlated with the VBM/CBM values calculated with respect to vacuum using DFT calculations of surface slabs. We can get useful predictions of the vacuum-aligned VBM by subtracting a heuristic constant value of 10 eV from the ALIGNN predictions, and adding the bulk TBmBJ band gap to get the corresponding CBM. } We visualize this IU based band alignment using ALIGNN model in Fig. \ref{fig:offsetexample}b. We observe that the overall trends of DFT and ALIGNN closely resemble each other. For these surfaces, we calculate the classification accuracy of the heterostructures in type-I, type-II and type-III heterostructures. We find precision scores of 66.7 \%, 66.1 \% and 58.2 \% respectively. Precision is defined as the fraction of relevant instances among all of the retrieved instances. The classification precision scores are based on DFT optimized surfaces, which are not available for all the materials in the database. So, we generate structures directly from the bulk counterparts, relax them using ALIGNN-FF and then predict the electron affinity and ionization potentials using the procedure mentioned above. In this way, we find precision scores of  55.0 \%, 63.4 \% and 60.0 \% respectively suggesting that structure optimization of surfaces has an impact on the ALIGNN predictions. \textcolor{black}{The random baseline is 1/3 = 33 \%, which is more than 2 times lower than what we achieve.} 

\textcolor{black}{As an example of the type of analysis that can be done with this data, we analyze all hetrostructures where the film is silicon. As shown in Fig. \ref{fig:ptable}, we identify which elements in the second semiconductor make it most likely that the hetrostructrue will have a Type-1/straddling band alignment appropriate for diode applications. We find these elements to be Al, P, S, N, O, Li which is consistent with known silicon devices. Many other analyses are possible, we provide this data in hopes that it will be useful to the community.}  

Sufficiently high precision scores suggest that such models can be used for pre-screening applications followed by density functional theory calculations and experiments. Also, as the DFT bulk, surface and interface dataset is growing continuously, there is plenty of scope to improve the model performance in the future. For the 1.4 trillion semiconductor interfaces, we find 294 billion as type-I, 322 billion as type-II and rest as type-III heterostructures using the ALIGNN+IU model. The results suggest that finding type-I interfaces for transistor applications is more challenging than other heterostructure applications. Having such a large number of options and further screened for desirable properties such as effective masses, dielectric, piezoelectric, thermoelectric properties etc. can be helpful for technological applications. We emphasize the point that AI models should be considered as a pre-screening step only and would require thorough DFT and/or experimental validation.

\textcolor{black}{Given the lack of surface-specific training data, the level of agreement with the unseen surface data is surprising, however, there are discrepancies in some cases. In other words, the model has never seen bonding environments that occur on slab surfaces, and extrapolating to such environments is challenging. The goal here is to obtain a fast model that can be used for quick screening of surfaces, with subsequent DFT calculations for confirmation. It may be possible to finetune this ALIGNN model with a surface dataset to further improve the accuracy of the model, but we leave that for future work. Nevertheless, the close resemblance in alignment predictions is promising and suggests that our models can be useful. Similar successful extrapolations for bulk-trained models were observed in Ref. \cite{choudhary2023can}, which demonstrates that vacancy energies can be predicted from a ML model fit to bulk crystal data only. } \textcolor{black}{We clarify that we are not using a ALIGNN model to predict either the OptB88vdW or TBmBJ band gaps in this work. We are simply looking up the bulk band gaps in the JARVIS-DFT database, so this does not contribute to the error. However, we do have models for these quantities with MAEs of 0.14 eV and 0.31 eV respectively (see Ref\cite{choudhary2021atomistic,choudhary2023large}). For materials not in the JARVIS-DFT database, we could use these models to predict the gaps.
}



In summary, we have provided a computational framework and dataset for investigating interface systems using multi-fidelity computational approaches. We have developed one of the largest datasets, containing \nsurfaces surfaces, \iuband IU-band offsets, and \nasj ASJ interface band offset using DFT. Using universal graph neural network models, we have quickly screened potential semiconductor device candidates as transistors from a pool of 1.4 trillion possible interfaces, which would not have been possible using conventional computational or experimental techniques. Although we have applied this framework for semiconductors, it can be useful for other technological applications as well. After pre-screening, we have shown and benchmarked this streamlined workflow for band offset predictions using the independent unit and alternate slab junction models. This work paves the way for the application of materials design approach to interface systems. All of the tools and datasets developed in this work will be distributed publicly in the spirit of open-science.

\section*{Methods}


\textcolor{black}{Graph neural networks are trained using Atomistic Line Graph Neural Network (ALIGNN) framework \cite{choudhary2021atomistic} which uses PyTorch and deep graph library (DGL). Such GNN models can be used for graph level prediction (such as total energy of the system, bandgap etc.) or node level predictions (such forces, charges, atomic magnetic moments etc.) In ALIGNN, a crystal structure is represented as a graph using atomic elements as nodes and atomic bonds as edges. Each node in the atomistic graph is assigned 9 input node features based on its atomic species: electronegativity, group number, covalent radius, valence electrons, first ionization energy, electron affinity, block and atomic volume. The inter-atomic bond distances are used as edge features with radial basis function up to 8 $\textrm{\AA}$ cut-off and a 12-nearest-neighbor ($N$). This atomistic graph is then used for constructing the corresponding line graph using interatomic bond-distances as nodes and bond-angles as edge features. ALIGNN uses edge-gated graph convolution for updating nodes as well as edge features using a propagation function ($f$) for layer ($l$), atom features ($h$), and node ($i$), details of which can be found in Ref. \cite{choudhary2021atomistic, choudhary2023unified}: }

\begin{equation} 
h_i^{(l+1)}=f(h_i^l{\{h_j^l}\}_{_i})
\end{equation}

ALIGNN is trained for 500 epochs and with default parameters in the package. We use a 90:5:5 training:validation:testing \textcolor{black}{randomly distributed data split for the CBM and VBM of the bulk materials dataset. The data splits and corresponding identifiers used during the training are made available in the figshare repository.} While ALIGNN was used as surrogate/property prediction model at a graph level, in the later version, we also included atomwise/nodewise property predictions such as forces. These forces are directly derived from the energies hence leading to force-field development (ALIGNN-FF). ALIGNN-FF was trained on the JARVIS-DFT dataset. \textcolor{black}{Note that, we did not need to modify the ALIGNN  model for surfaces because these GNN are based on the local environment around each atom only. We have used the same ALIGNN model in molecules \cite{choudhary2021atomistic} and metal-organic frameworks \cite{choudhary2022graph} also without changing any architecture and still leading to accurate results. Next, JARVIS-DFT is a collection of 80000 diverse materials primarily using OptB88vdW in VASP software following strict protocols for convergence etc. In addition to the datasets, JARVIS-DFT is seamlessly integrated with the JARVIS-tools package for setting up calculations and performing analysis using a variety of multi-fidelity and multi-scale simulation approaches.} ALIGNN-FF was trained on 307811 bulk structures with 1 million forces obtained from SCF relaxation step for materials in the JARVIS-DFT. ALIGNN-FF was shown to capture both structural and chemical diversity with reasonable accuracy especially for structure optimization. 
DFT calculations were carried out using the Vienna Ab-initio simulation package (VASP) software \cite{kresse1996efficient,kresse1996efficiency} with OptB88vdW \cite{klimevs2009chemical}, TBmBJ\cite{tran2009accurate} and R2SCAN \cite{furness2020accurate} functionals using the workflow given on our ‘jarvis-tools’ GitHub page (https://github.com/usnistgov/jarvis ). We use the OptB88vdW functional, which gives accurate lattice parameters for both vdW and non-vdW (3D-bulk) solids. The crystal structure was optimized until the forces on the ions were less than 0.01 eV/Å and energy less than $10^{-6}$ eV. Also, we calculate the local potential containing ionic plus Hartree contributions to determine the vacuum potential (VAC) of surface slabs. The VAC is subtracted from the valence band maxima (VBM) and conduction band minima (CBM) to enable the comparison of band-diagrams of individual slabs in band-alignment diagrams. The converged k-points and cut-off for the bulk materials were also used for the corresponding surface slab models. The ASJ based interface structures were generated using Zur algorithm \cite{zur1984lattice} available in the JARVIS-Tools. For a quick scan of xy-displacements for surfaces in the interfaces, ALIGNN-FF was used.


\section*{Acknowledgements}

We thank computational resource from National Institute of Standards and Technology (NIST).This work was performed with funding from the CHIPS Metrology Program, part of CHIPS for America, National Institute of Standards and Technology, U.S. Department of Commerce. Please note commercial software is identified to specify procedures. Such identification does not imply recommendation by National Institute of Standards and Technology (NIST).

\section*{Author contributions}

K.C. conceived the high-throughput workflow and conducted all calculations,  K.F.G helped in setting up calculations and analysing the results. All authors reviewed the manuscript. 
\section*{Competing interests}

The authors declare no competing interests.

\section*{Data availability}

The data generated by this work will be made publicly available at JARVIS websites: \url{https://pages.nist.gov/jarvis/databases}, \url{https://jarvis.nist.gov/jarvisdft/} and Figshare (\url{https://doi.org/10.6084/m9.figshare.25514719}, \url{https://doi.org/10.6084/m9.figshare.25832614}). A webapp will also be made available at the JAVRIS-Heterostructure website (\url{https://jarvis.nist.gov/jarvish/}).

\section*{Code availability}

The code used in this work, InterMat is made publicly available at : \url{https://github.com/usnistgov/intermat}. It depends on closely related codes available at \url{https://github.com/usnistgov/jarvis} and \url{https://github.com/usnistgov/alignn}.

\section*{Conflicts of interest}
There are no conflicts to declare.





\scriptsize{
\bibliography{rsc} 
\bibliographystyle{rsc} } 

\end{document}